
\input epsf



 \font\bigrm=cmr12 scaled\magstep3
 \font\twelverm=cmr12     \font\twelvei=cmmi12
 \font\twelvesy=cmsy10 scaled 1200  \font\twelveex=cmex10 scaled 1200
 \font\twelvebf=cmbx12   \font\twelvesl=cmsl12
 \font\twelvett=cmtt12  \font\twelveit=cmti12

 \skewchar\twelvei='177   \skewchar\twelvesy='60

\catcode`@=11 

\def\twelvepoint{\normalbaselineskip=12.4pt
   \abovedisplayskip 12.4pt plus 3pt minus 9pt
   \belowdisplayskip 12.4pt plus 3pt minus 9pt
   \abovedisplayshortskip 0pt plus 3pt
   \belowdisplayshortskip 7.2pt plus 3pt minus 4pt
   \smallskipamount=3.6pt plus1.2pt minus1.2pt
   \medskipamount=7.2pt plus2.4pt minus2.4pt
   \bigskipamount=14.4pt plus4.8pt minus4.8pt
   \def\rm{\fam0\twelverm}          \def\it{\fam\itfam\twelveit}%
   \def\sl{\fam\slfam\twelvesl}     \def\bf{\fam\bffam\twelvebf}%
   \def\mit{\fam 1}                 \def\cal{\fam 2}%
   \def\tt{\twelvett}
   \textfont0=\twelverm   \scriptfont0=\tenrm   \scriptscriptfont0=\sevenrm
   \textfont1=\twelvei    \scriptfont1=\teni    \scriptscriptfont1=\seveni
   \textfont2=\twelvesy   \scriptfont2=\tensy   \scriptscriptfont2=\sevensy
   \textfont3=\twelveex   \scriptfont3=\twelveex  \scriptscriptfont3=\twelveex
   \textfont\itfam=\twelveit
   \textfont\slfam=\twelvesl
   \textfont\bffam=\twelvebf \scriptfont\bffam=\tenbf
   \scriptscriptfont\bffam=\sevenbf
   \normalbaselines\rm}

\def\oneandahalfspace{\baselineskip=\normalbaselineskip
   \multiply\baselineskip by 3 \divide\baselineskip by 2}

\def\raggedcenter{\leftskip=1.5em plus 12em \rightskip=\leftskip
   \parindent=0pt \parfillskip=0pt \spaceskip=.3333em \xspaceskip=.5em
   \pretolerance=9999 \tolerance=9999
   \hyphenpenalty=9999 \exhyphenpenalty=9999 }

\newcount\figno\global\figno=0
\newcount\tabno\global\tabno=0

\def\fig#1#2{
	\global\advance\figno by 1
	\expandafter\xdef\csname f@g#1\endcsname{\t@ghead\number\figno}
	{\par\raggedcenter Fig.\t@ghead\number\figno{}. #2\smallskip}
	\immediate\write3{\string\figref{#1}{\t@ghead\number\figno}}
}

\def\rfig#1{\expandafter\ifx\csname f@g#1\endcsname\relax
\hbox{\it Fig. ??}
\message{Missing Figure reference (#1) -- may need to ReTeX}
\else\hbox{{Fig.}\csname f@g#1\endcsname}\fi}

\def\figref#1#2{\expandafter\xdef\csname f@g#1\endcsname{#2}}

\def\table#1#2{
	\global\advance\tabno by 1
	\expandafter\xdef\csname t@b#1\endcsname{\t@ghead\number\tabno}
	{\raggedcenter Table \t@ghead\number\tabno{}. #2\smallskip}
	\immediate\write3{\string\tablref{#1}{\t@ghead\number\tabno}}
}

\def\rtable#1{\expandafter\ifx\csname t@b#1\endcsname\relax
\hbox{\it Table ??}
\message{Missing Table reference (#1) -- may need to ReTeX}
\else\hbox{{Table} \csname t@b#1\endcsname}\fi}

\def\tablref#1#2{\expandafter\xdef\csname t@b#1\endcsname{#2}}

\def\tag#1$${\eqno(#1)$$}
\newcount\tagnumber\tagnumber=0

\immediate\newwrite\eqnfile
\newif\if@qnfile\@qnfilefalse
\def\write@qn#1{}
\def\writenew@qn#1{}
\def\w@rnwrite#1{\write@qn{#1}\message{#1}}
\def\@rrwrite#1{\write@qn{#1}\errmessage{#1}}

\def\t@ghead{}

\expandafter\def\csname @qnnum-3\endcsname
  {{\t@ghead\advance\tagnumber by -3\relax\number\tagnumber}}
\expandafter\def\csname @qnnum-2\endcsname
  {{\t@ghead\advance\tagnumber by -2\relax\number\tagnumber}}
\expandafter\def\csname @qnnum-1\endcsname
  {{\t@ghead\advance\tagnumber by -1\relax\number\tagnumber}}
\expandafter\def\csname @qnnum0\endcsname
  {\t@ghead\number\tagnumber}
\expandafter\def\csname @qnnum+1\endcsname
  {{\t@ghead\advance\tagnumber by 1\relax\number\tagnumber}}
\expandafter\def\csname @qnnum+2\endcsname
  {{\t@ghead\advance\tagnumber by 2\relax\number\tagnumber}}
\expandafter\def\csname @qnnum+3\endcsname
  {{\t@ghead\advance\tagnumber by 3\relax\number\tagnumber}}

\def\callall#1{\xdef#1##1{#1{\noexpand\call{##1}}}}
\def\call#1{\each@rg\callr@nge{#1}}
\def\(#1){(\call{#1})}

\def\each@rg#1#2{{\let\thecsname=#1\expandafter\first@rg#2,\end,}}
\def\first@rg#1,{\thecsname{#1}\apply@rg}
\def\apply@rg#1,{\ifx\end#1\let\next=\relax%
\else,\thecsname{#1}\let\next=\apply@rg\fi\next}

\def\callr@nge#1{\calldor@nge#1-\end-}
\def\callr@ngeat#1\end-{#1}
\def\calldor@nge#1-#2-{\ifx\end#2\@qneatspace#1 %
  \else\calll@@p{#1}{#2}\callr@ngeat\fi}
\def\calll@@p#1#2{\ifnum#1>#2{\@rrwrite{Equation range #1-#2\space is bad.}
\errhelp{If you call a series of equations by the notation M-N, then M and
N must be integers, and N must be greater than or equal to M.}}\else%
 {\count0=#1\count1=#2\advance\count1
by1\relax\expandafter\@qncall\the\count0,%
  \loop\advance\count0 by1\relax%
    \ifnum\count0<\count1,\expandafter\@qncall\the\count0,%
  \repeat}\fi}

\def\@qneatspace#1#2 {\@qncall#1#2,}
\def\@qncall#1,{\ifunc@lled{#1}{\def\next{#1}\ifx\next\empty\else
  \w@rnwrite{Equation number \noexpand\(>>#1<<) has not been defined yet.}
  >>#1<<\fi}\else\csname @qnnum#1\endcsname\fi}

\let\eqnono=\eqno
\def\eqno(#1){\tag#1}
\def\tag#1$${\eqnono(\displayt@g#1 )$$}

\def\aligntag#1\endaligntag
  $${\gdef\tag##1\\{&(##1 )\cr}\eqalignno{#1\\}$$
  \gdef\tag##1$${\eqnono(\displayt@g##1 )$$}}

\def\eqalignno#1{\displ@y \tabskip\centering
  \halign to\displaywidth{\hfil$\displaystyle{##}$\tabskip\z@skip
    &$\displaystyle{{}##}$\hfil\tabskip\centering
    &\llap{$\displayt@gpar##$}\tabskip\z@skip\crcr
    #1\crcr}}

\def\displayt@gpar(#1){(\displayt@g#1 )}

\def\displayt@g#1 {\rm\ifunc@lled{#1}\global\advance\tagnumber by1
        {\def\next{#1}\ifx\next\empty\else\expandafter
        \xdef\csname @qnnum#1\endcsname{\t@ghead\number\tagnumber}\fi}%
  \writenew@qn{#1}\t@ghead\number\tagnumber\else
        {\edef\next{\t@ghead\number\tagnumber}%
        \expandafter\ifx\csname @qnnum#1\endcsname\next\else
        \w@rnwrite{Equation \noexpand\tag{#1} is a duplicate number.}\fi}%
  \csname @qnnum#1\endcsname\fi}

\def\ifunc@lled#1{\expandafter\ifx\csname @qnnum#1\endcsname\relax}

\let\@qnend=\end\gdef\end{\if@qnfile
\immediate\write16{Equation numbers written on []\jobname.EQN.}\fi\@qnend}

\def\endpage{\vfill\eject}

\def\pagenumbers{\global\pageno=1\global\footline={\hfil\rm\folio\hfil}}

\def\lsim{\mathrel{\mathpalette\@versim<}}
\def\gsim{\mathrel{\mathpalette\@versim>}}
\def\@versim#1#2{\lower0.2ex\vbox{\baselineskip\z@skip\lineskip\z@skip
  \lineskiplimit\z@\ialign{$\m@th#1\hfil##\hfil$\crcr#2\crcr\sim\crcr}}}

\catcode`@=12 

\openin4=\jobname.lab \ifeof4\else\closein4\input \jobname.lab\fi
\immediate\openout 3=\jobname.lab

\def\head#1{\goodbreak\bigskip\leftline{\bf #1}}

\def\ie{{\it i.e., }}
\def\eg{{\it e.g. }}
\def\etal/{{\it et al.}}
\def\fb{\hbox{ fb}}
\def\GeV{\hbox{ GeV}}
\def\TeV{\hbox{ TeV}}
\def\MeV{\hbox{ MeV}}
\def\LEPI/{\hbox{LEP I}}
\def\LEPII/{\hbox{LEP I$\!$I}}
\def\Zstar{Z^{\scriptscriptstyle ( \scriptstyle * \scriptscriptstyle )}}
\def\gammastar{\gamma^{\scriptscriptstyle ( \scriptstyle *
\scriptscriptstyle )}}
\def\Br{\hbox{Br}}
\def\pb{\hbox{pb}}

\twelvepoint
\oneandahalfspace

\hoffset=0cm
\voffset=0cm
\hsize=15.5 true cm
\vsize=22 true cm

\def\refradmt{[1]}
\def\refHHG{[2]}
\def\refHmass{[3]}
\def\refWu{[4]}
\def\refHZZ{[5]}
\def\refHBr{[6]}
\def\refHff{[7]}
\def\refffake{[8]}
\def\refWH{[9]}
\def\refttH{[10]}
\def\refttHbb{[11]}
\def\refMRS{[12]}
\def\refjetfake{[13]}
\def\refmettff{[14]}
\def\refttff{[15]}
\def\refmeqqttH{[16]}
\def\refNLOWf{[17]}
\def\refradzero{[18]}
\def\refradzeroII{[19]}
\def\refWffj{[20]}
%
%
%

\nopagenumbers

\hfill\vbox{\hbox{MAD/PH/796}
\hbox{October 1993}}
\vskip 1cm
{\raggedcenter
\bigrm The Search for a Light `Intermediate Mass' Higgs Boson.

}
\vskip 1cm
{\raggedcenter
\bf D.J.Summers

}

\vskip 0.7cm {\raggedcenter\obeylines\sl
Department of Physics,
University of Wisconsin -- Madison,
1150 University Avenue,
Madison,
WI 53706,
U.S.A.
}

\vskip 3cm

\centerline{\bf Abstract}
{We review recent progress in techniques for searching for the Standard Model
light `Intermediate Mass' Higgs Boson ($80 \GeV \lsim M_H \lsim 140\GeV$).
We pay particular attention to associated production at the SSC and LHC
where we search for the Higgs produced in association
with either a $W$ boson or
a $t$ quark. This production mode can be detected cleanly when the Higgs
decays into two isolated photons, and the associated heavy particle decays
semi-leptonically. We discuss the possibility that radiative corrections may
significantly  enhance the $W \gamma\gamma$ background, decreasing the
significance of the  $WH$ signal.}

\vskip 1cm
{\sl Based upon a Plenary talk given at the SSC Symposium, Madison,
March 1993.}

\endpage
\pagenumbers
\head{Introduction}

The Standard Model of Particle Physics, based upon a spontaneously broken
$SU(3) \otimes SU(2) \otimes U(1)$
gauge theory, has proven to be remarkably
successful over the past 20 years. However two major predictions of this model
are the existence of a sixth quark, the top quark, and of a neutral $CP$-even
scalar particle, the Higgs boson; neither of which have so far been observed.
Through higher order corrections to observed processes we expect that  $m_t
\lsim 200 \GeV$ \refradmt,
and so should be detected at the Tevatron at Fermilab
within the next few years. On the other hand the Higgs boson has far fewer
constraints placed upon it. Theoretically, both from perturbative and lattice
calculations, we expect $M_H \lsim 1 \TeV$ \refHHG. Direct searches at \LEPI/
give $M_H > 62.5 \GeV$ at the 95\% confidence level \refHmass.

At \LEPII/ we expect the Standard Model Higgs mass bound to be raised to
$80\GeV$ \refWu, however due to the limited centre of mass energy  ($\sqrt s
\lsim 180 \GeV$) we are unlikely to be sensitive to Higgs with mass larger than
$80\GeV$. The hadron super colliders the SSC and the LHC will be able to detect
a heavy Higgs with $M_H \gsim 140 \GeV$ \refHZZ, such a Higgs
has an appreciable branching ratio into $\Zstar\Zstar$ (where the $Z$
bosons are either on or off mass shell), and hence into four
leptons ($H \to  \Zstar\Zstar \to 4l$ with $l=e,\mu$) (see \rfig{Hbr}). In this
$4l$ decay mode a heavy Higgs can be detected cleanly. For masses
beneath $M_H\approx 130\GeV$ the $\Zstar\Zstar$ branching ratio falls
rapidly, this means that the $4l$
signal falls beneath irreducible backgrounds from  $\Zstar\Zstar$ and  $\Zstar
\gammastar$ production, and reducible backgrounds from $t \bar t$ and $b \bar b
Z$ production \refHZZ.
\topinsert
\centerline{\epsfxsize=9 cm\epsfbox{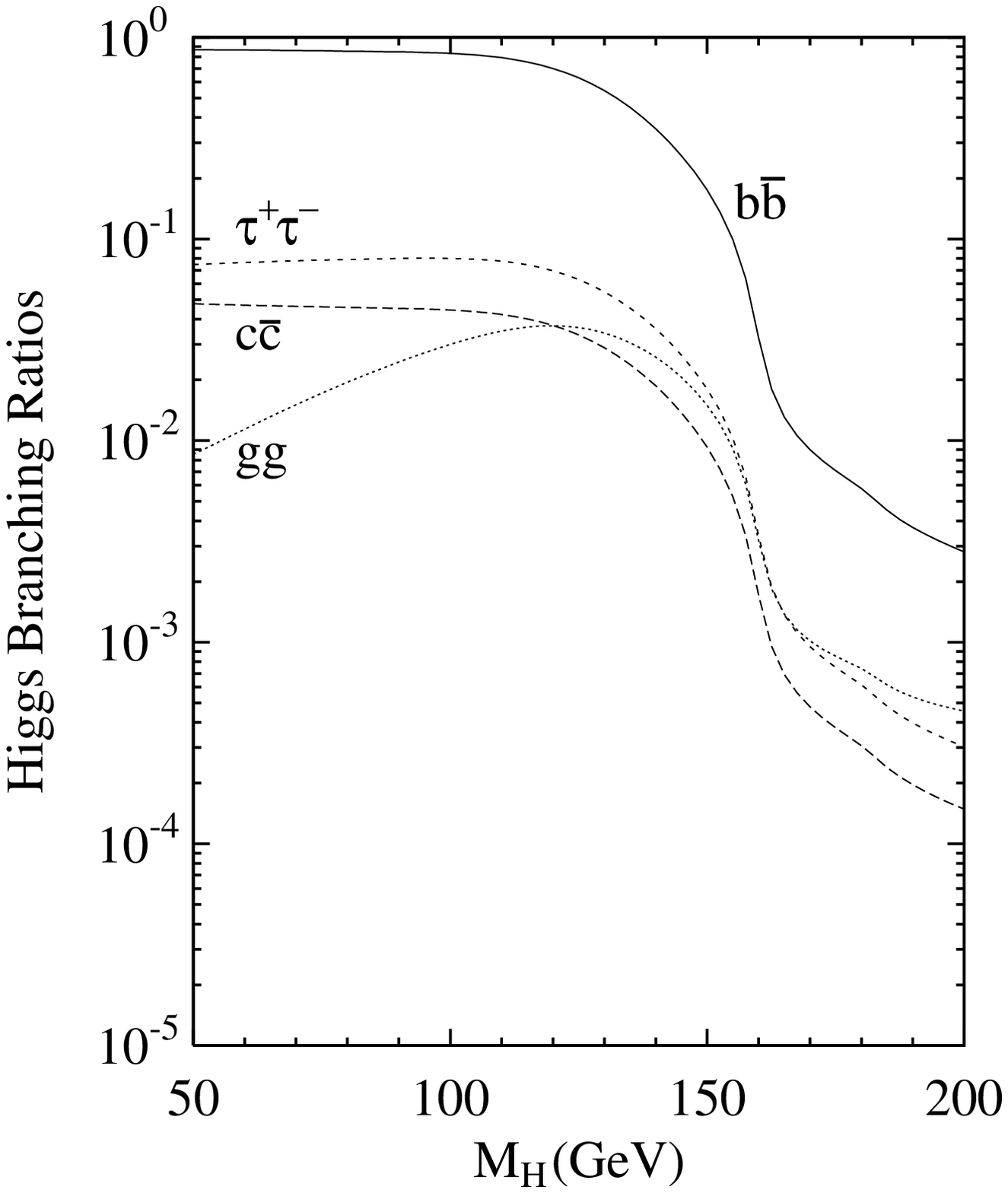}
\qquad\epsfxsize=9 cm\epsfbox{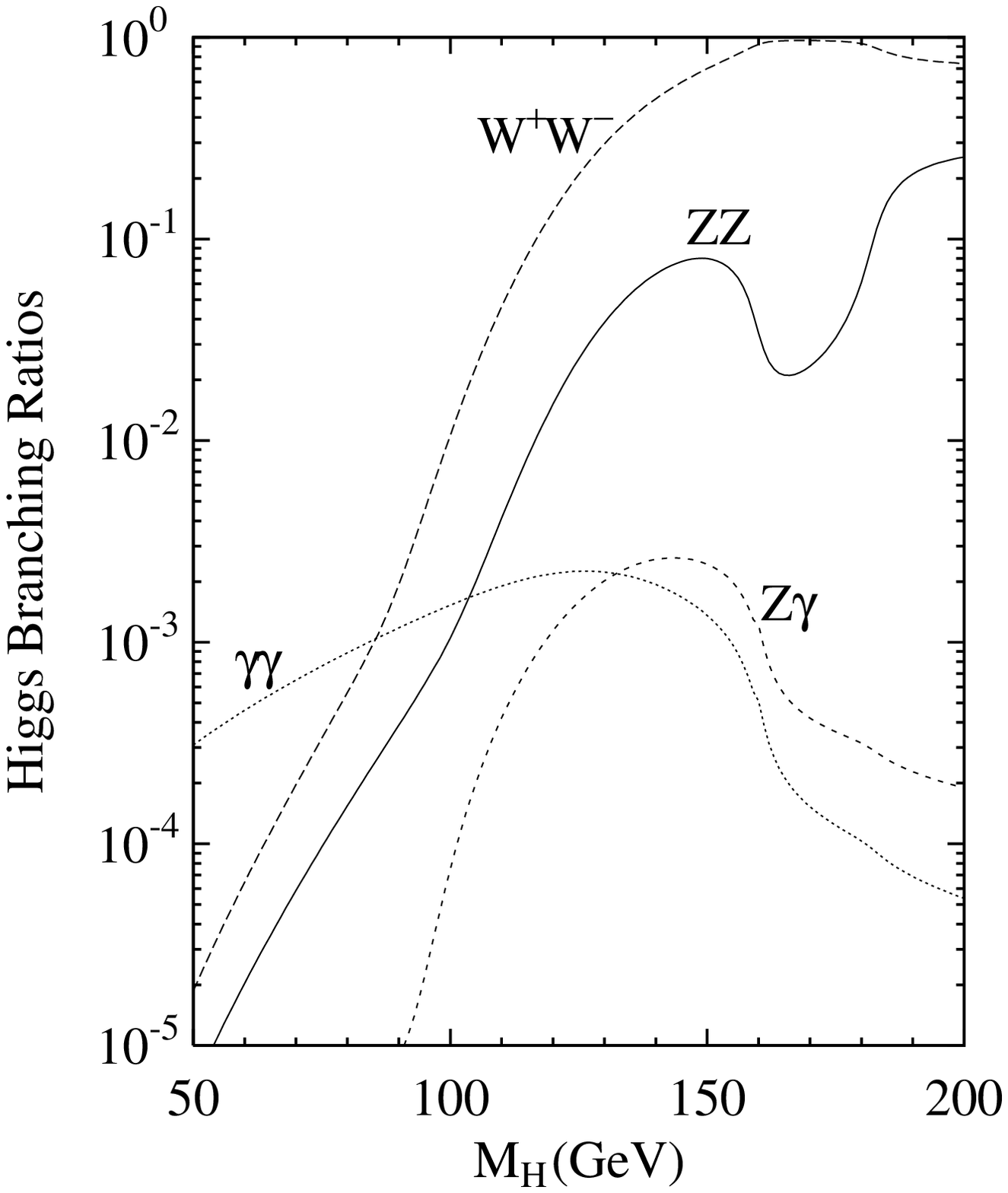}}
\fig{Hbr}{Branching Ratios of an `intermediate mass' Standard Model Higgs
boson \refHBr.}
\endinsert

For masses beneath $M_H \approx 140 \GeV$ the major Higgs decay mode into jets
are swamped by vast QCD backgrounds, this leaves only the modest $H \to
\gamma\gamma$ decay mode as remotely observable \refHff. Nonetheless despite
the modest branching ratio into 2 photons
${\cal O} (10^{-3})$, see \rfig{Hbr},
we have numerous events, because for a light Higgs we expect large
numbers of Higgs bosons to be produced by gluon--gluon
annihilation via a top quark loop \refHHG.
Unfortunately as well as numerous events, we
also have  huge backgrounds: both reducible from jets faking photons, and
irreducible from $q \bar q \to \gamma \gamma$ and  $gg \to \gamma \gamma$
production. The reducible background can be reduced to an insignificant level
if we can distinguish jets from photons to 1 part in $10^4$ \refffake ;
however the continuum $q \bar q \to \gamma \gamma$ and  $gg \to \gamma \gamma$
irreducible background processes are by themselves huge.
This places severe constraints upon our resolution for the
measurement of the diphoton invariant mass; in practice we require this
measurement to be accurate to about the $1 \%$ level, which will be very hard
to achieve with a general purpose detector. Even
with such an excellent detector
we still have a low signal to background ratio, but due to the large number of
events we can still get a high significance. Typically there are  ${\cal
O}(1000)$ signal events on top of a continuum background of  ${\cal O}(10000)$
background events, which would give an  ${\cal O}(10)$ s.d. effect.

Clearly this makes the light `intermediate mass' Higgs with
$ 80 \GeV \lsim M_H \lsim 140 \GeV$,  exceptionally hard to detect.

\head{Associated Production}

In the past few years an alternative approach to looking for the the light
`intermediate mass' Higgs at proton--proton colliders has been proposed
\refWH \refttH . Rather than looking for the Higgs in isolation we look for it
produced in association with other particles. Since the Standard Model Higgs is
responsible for mass generation it couples predominantly
to heavy particles (\ie
$W$'s, $Z$'s, and $t$ quarks), and so if we have an event that is tagged as
containing a heavy particle there is a far greater probability that it will
also contain a Higgs boson relative to a typical event.
By searching for a Higgs
plus a heavy particle we hope to achieve far better signal to background
ratios. Of course by using such a strategy we expect a lower signal rate due
to the extra particle produced, but hope that the gain in the signal to
background ratio compensates for the lower event rate. We should not forget
that for isolated Higgs production it is not for lack of signal that the Higgs
is hard to detect, but that huge backgrounds overwhelm our signal.

There are 3 cases associated with the massive particles, the $Z$ and $W$
bosons, and the $t$ quark. The dominant decays of the Higgs are swamped
by huge backgrounds (although there has been some recent interest in  $pp \to t
\bar t H \to t \bar t b \bar b$ \refttHbb), and again we are forced to the
Higgs decay $H \to \gamma\gamma$.

\head{$ZH$ Associated Production}

In the process $pp \to ZH$ we can cleanly tag the $Z$ in its decay,
$Z \to l^+ l^- $ for $l=e,\mu$. Unfortunately this process has a
branching ration of only 7\% and this means
that there are too few events to be
detected \refWH . This is true even at a high luminosity SSC with
${\cal L}= 10^5 \hbox{pb}^{-1}$.

\head{$WH$ Associated Production}

In the process $pp \to WH$ the $W$ boson can be cleanly tagged in its leptonic
decay modes, $W \to l \nu$ for $l=e,\mu$, although we only have a single
lepton. This process is similar to $ZH$ production, however the production rate
is about 6 times larger than $ZH$ production due to a larger
leptonic branching ratio
$\Br ( W \to l \nu ) = 20 \%$ (c.f. $\Br (Z \to l^+l^-) = 7\%$) and also
because  the $V\!-\!A$ couplings of the $W$ to the initial state quarks are
larger than the corresponding $Z$ couplings \refWH .
%
%
Concentrating on the $pp \to WH \to l \gamma \gamma$ signal, the lowest order
subprocess is
$$
q \bar q' \to WH \to l \nu \gamma \gamma \qquad .\tag WH
$$
For the lepton (from the $W$ decay) and
the photons (from the Higgs decay) to be detected we insist that they pass the
cuts,
$$
p_\perp > 20\GeV \qquad , \qquad | \eta | < 2.5 \qquad .\tag cuts
$$
Now if we allow the photons to become too close to any hadronic activity then
we have a large background from $\pi^0$'s faking photons; also if the
lepton becomes close to hadronic activity we have a large background from
semileptonic $b$ and $c$ quark decays.
Further as the photons become collinear to
the lepton the backgrounds have a collinear singularity coming from photons
being radiated from the lepton. To remove these backgrounds we need to
insist that the photons and leptons are well separated from any jets and from
each other. This is implemented by insisting that the leptons and photons are
separated from jets and each other by a minimum $\Delta R$;
$$
\Delta R > 0.4 \qquad \hbox{where }
\Delta R = \sqrt{ \Delta \phi^2 + \Delta \eta^2} \qquad .\tag DeltaR
$$
\topinsert
\centerline{
\vbox{\epsfxsize=9 cm\epsfbox{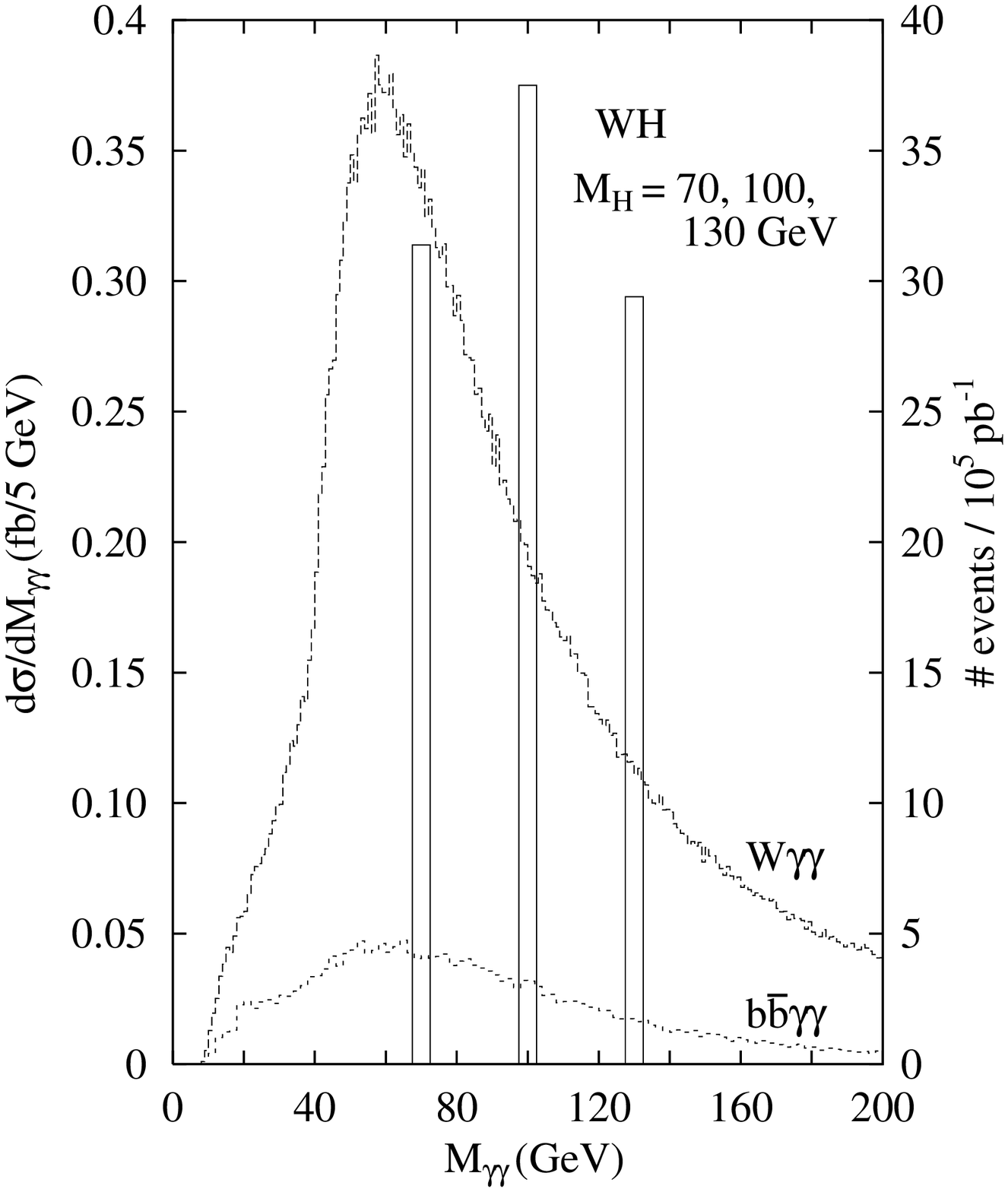}
\hbox{\hskip 1.5cm a) At the LHC with  $\sqrt s = 16 \TeV$}}
\qquad
\vbox{\epsfxsize=9 cm\epsfbox{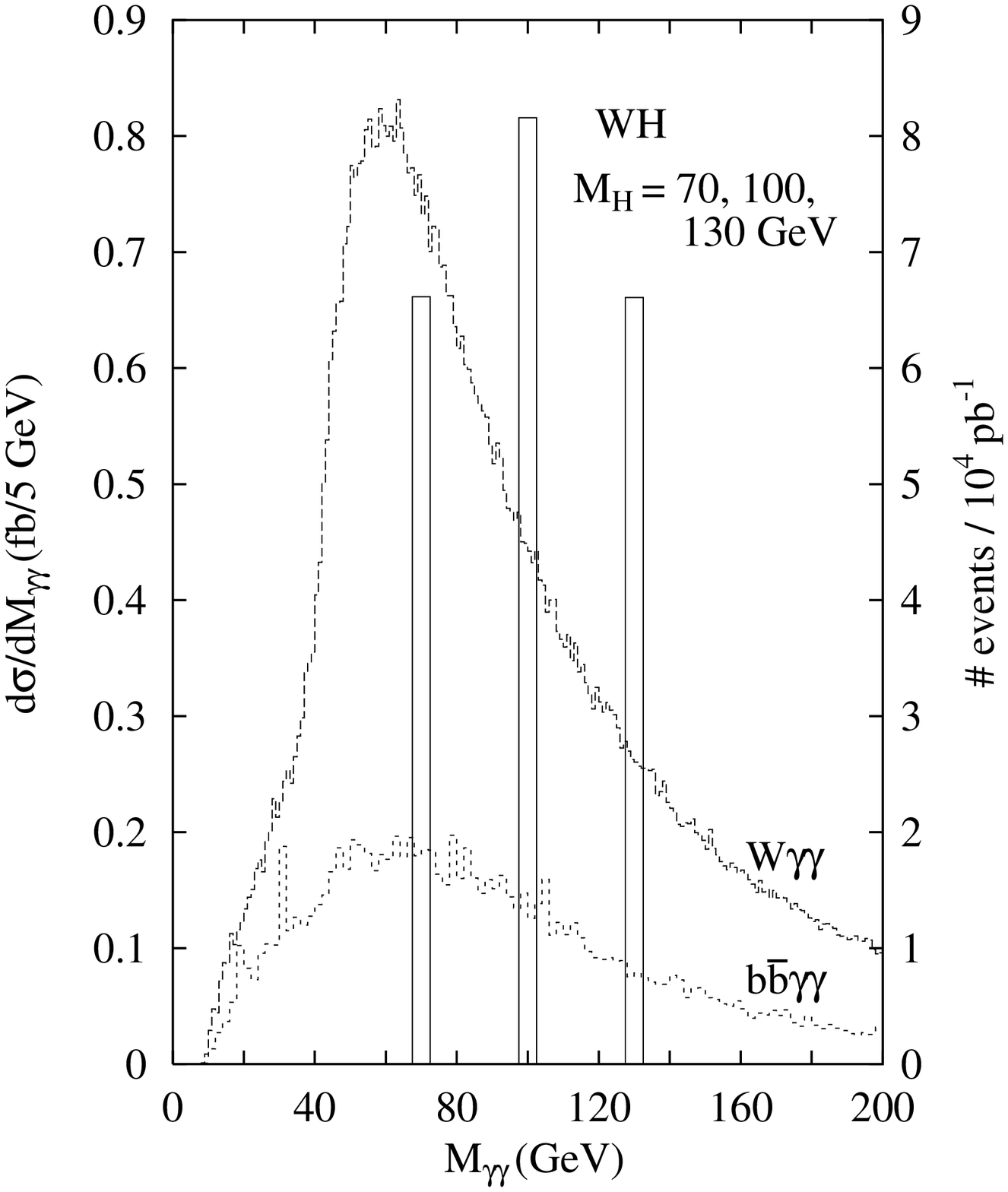}
\hbox{\hskip 1.5cm b) At the SSC with  $\sqrt s = 40 \TeV$}}}
\fig{WHcross}{The cross-section for the signal
$pp \to WH \to l \gamma \gamma$, the main irreducible background,
$pp \to W\gamma\gamma \to l \gamma \gamma$, and reducible background,
$pp \to b \bar b \gamma\gamma \to l \gamma \gamma$. Also shown are
the expected number of events at the high luminosity LHC with
${\cal L}=10^5 \pb^{-1}$ and the standard luminosity SSC with
${\cal L}=10^4 \pb^{-1}$.}
\endinsert
\noindent
With these cuts the dominant background is the irreducible
background \refWH,
$$
pp \to q \bar q ' \to W \gamma \gamma \qquad .\tag Wff
$$
The cross-sections and events rates for the signal \(WH) and the background
\(Wff) are shown in \rfig{WHcross}. We use the MRS92$'$ D0 set of parton
distributions \refMRS , and for consistency
with these parton distributions choose
$\Lambda_{\overline{MS}}(\hbox{nf}=4)=230 \MeV$, which we have rescaled to
a five flavour $\Lambda_{\overline{MS}}$.

If we can achieve a rejection of jets faking photons to 1 part in $10^4$, \ie
the same rejection factor that would be required if we are to detect the Higgs
in its two photon decay mode, then we can reduce to an insignificant level
the background from \refjetfake ,
$$
\eqalign{
pp &\to W \hbox{\it jet jet} \to W \gamma \gamma \cr
pp &\to W \gamma\hbox{ \it jet} \to W \gamma \gamma \qquad .\cr
}\tag
$$
As well as the irreducible backgrounds we also have reducible backgrounds, in
particular coming from heavy quark decays faking isolated leptons \refWH
\refmettff ,
$$
pp \to b \bar b \gamma\gamma \to l \gamma \gamma X \tag
$$
Conservatively we expect that only 1/7 semileptonic $b$ decays will pass our
isolation cut \(DeltaR) \refHff . Using this as the rejection factor
coming from isolation, and calculating the dominant
gluon fusion subprocess \refmettff ,
$$
pp \to gg \to b \bar b \gamma\gamma \to l \gamma \gamma X \tag
$$
which we expect to dominate at the LHC and SSC,
we se that this process does not
present a serious background. The rate is show in \rfig{WHcross}. $c$ quark
decays will very rarely produce an {\it isolated} lepton, and so given the $b$
quarks decays to leptons are not a significant background to $WH$ detection we
do not expect $c$ quark decays to pose a problem.

In \rfig{WHcross} we have chosen $5\GeV$ bins for
the photon--photon invariant mass; in practice we
expect to be able to achieve a far better experimental resolution than
this. Binning the cross-section in narrower bins will cause the
continuum backgrounds to be lowered, without
lowering the signal at all (remember that for a `intermediate mass'
Higgs its width is ${\cal O}(\MeV)$ and so is far smaller than any
experimental resolution).
It is clear from \rfig{WHcross} that we have good signal
to background ratios, far better than the signal to background ratios
for the plain $H \to \gamma \gamma$ detection mode for the Higgs.
However our signal rate is low, we have only a handful of events; this
means that we will require high luminosity at both the
LHC and the SSC. In \rtable{WHcross} we show the numbers of signal events
expected at the high luminosity LHC and SSC with ${\cal L}=10^5 \pb^{-1}$
we also show the numbers of background events for 3\%
$M_{\gamma\gamma}$ resolution, and give the significance of this signal
if it were to be observed.
\topinsert
\centerline{\vbox{\offinterlineskip
\hrule
\halign{\vrule#&\strut\quad\hfil$#$\quad
&\vrule#&\quad\hfil$#$\quad
&\vrule#&\quad\hfil$#$\quad
&\vrule#&\quad\hfil$#$\quad
&\vrule#&\quad\hfil$#$\quad
&\vrule#&\quad\hfil$#$\quad
&\vrule#&\quad\hfil$#$\hfil\quad
&\vrule#\cr
height2pt&\omit&&\omit&&\omit&&\omit&&\omit&&\omit&&\omit&\cr
&\sqrt s\hfil&&m_H\hfil&& \Delta m\hfil && WH\hfil && W\gamma\gamma\hfil &&
 b \bar b \gamma\gamma\hfil && $Significance $ &\cr
&(\TeV)&&(\GeV)&& (\GeV)&& $\# events$  && $\# events$ && $\# events$ &&
= {N_{\rm Sig} \over \sqrt{N_{\rm Back}}} &\cr
height2pt&\omit&&\omit&&\omit&&\omit&&\omit&&\omit&&\omit&\cr
\noalign{\hrule}
height2pt&\omit&&\omit&&\omit&&\omit&&\omit&&\omit&&\omit&\cr
&     && 60  &&  1.8 && 27.6 &&  13.4 && 1.6 && 7.1  &\cr
&     && 70  &&  2.1 && 32.9 &&  14.1 && 1.7 && 8.3  &\cr
&     && 80  &&  2.4 && 36.2 &&  14.0 && 1.8 && 9.1  &\cr
&     && 90  &&  2.7 && 38.2 &&  12.8 && 1.9 && 10.0 &\cr
&     && 100 &&  3.0 && 39.2 &&  11.7 && 1.9 && 10.6 &\cr
& 16  && 110 &&  3.3 && 38.8 &&  10.9 && 1.6 && 11.0 &\cr
&     && 120 &&  3.6 && 35.7 &&   9.6 && 1.5 && 10.7 &\cr
&     && 130 &&  3.9 && 29.4 &&   8.9 && 1.3 && 9.2  &\cr
&     && 140 &&  4.2 && 20.8 &&   8.1 && 1.0 && 6.9  &\cr
&     && 150 &&  4.5 && 12.0 &&   7.3 && 1.0 && 4.2  &\cr
&     && 160 &&  4.8 &&  3.8 &&   6.7 && 0.9 && 1.4  &\cr
height2pt&\omit&&\omit&&\omit&&\omit&&\omit&&\omit&&\omit&\cr
\noalign{\hrule}
height2pt&\omit&&\omit&&\omit&&\omit&&\omit&&\omit&&\omit&\cr
&     && 60  &&  1.8 && 57.8 &&  29.2 && 6.4 &&  9.7 &\cr
&     && 70  &&  2.1 && 69.3 &&  31.4 && 7.6 && 11.1 &\cr
&     && 80  &&  2.4 && 77.0 &&  30.4 && 8.0 && 12.4 &\cr
&     && 90  &&  2.7 && 82.1 &&  28.9 && 8.4 && 13.4 &\cr
&     && 100 &&  3.0 && 85.3 &&  26.6 && 8.0 && 14.5 &\cr
& 40  && 110 &&  3.3 && 85.3 &&  24.2 && 7.6 && 15.1 &\cr
&     && 120 &&  3.6 && 79.5 &&  22.3 && 6.5 && 14.8 &\cr
&     && 130 &&  3.9 && 66.1 &&  20.5 && 5.9 && 12.9 &\cr
&     && 140 &&  4.2 && 47.2 &&  18.7 && 6.0 &&  9.5 &\cr
&     && 150 &&  4.5 && 27.7 &&  17.2 && 5.4 &&  5.8 &\cr
&     && 160 &&  4.8 &&  8.8 &&  16.0 && 4.8 &&  1.9 &\cr
height2pt&\omit&&\omit&&\omit&&\omit&&\omit&&\omit&&\omit&\cr}
\hrule}}
\table{WHcross}{The expected numbers of $pp \to WH\to l\gamma\gamma$
signal events and background events at the
high luminosity LHC and SSC (${\cal L}=10^5 \pb^{-1}$) for various
mass values  of the Standard Model Higgs boson assuming reasonable
diphoton mass resolution of about 3\%. We also give the significance
in s.d. that this signal would be if observed.}
\endinsert

\head{$t\bar t H$ Associated production}

Moving on to $pp \to t \bar t H$ production \refttH\refmettff\refttff;
within the Standard Model 100\% of
$t$ quarks decay to $W$ bosons; and so as with $W$ detection we can tag the
heavy particle on an isolated lepton. This means that the signal that we look
for is again an isolated lepton, and two isolated photons. Despite the fact
that our signal is the same as for $WH$ production, the two processes look
considerable different. At leading order $WH$
production just has an observed lepton and
two photons in the final state; whereas $t\bar tH$ production in addition a
lepton and two photons also typically has 4 jets (1 jet coming from the
semileptonic $t$ decay, and 3 jets coming from the other $t$ decay). This
makes the $t\bar tH$ final state far more active than the $WH$ final state,
and in particular means that
the backgrounds for $WH$ and $t \bar t H$ production are distinct. The $WH$
cross-section is proportional to the $WWH$ coupling,
and the $t \bar t H$ process
is proportional to the $ttH$ coupling; by measuring these two processes
independently we can separately extract the $WWH$ and $ttH$ couplings.

We apply the same cuts as in the $WH$ production case \(cuts,DeltaR). As well
as an isolated lepton from the $t$ decay
we gain a factor of 2 from leptons coming from the $\bar t$
decay. We calculate both the $gg \to t \bar t H$ and  $q \bar
q \to t \bar t H$ cross-sections; due to the large gluon luminosity at the LHC
and SSC the $gg$ fusion process dominates, however the $q \bar q$ initial state
contributes 10\% at the SSC and  25\% at the LHC of the $gg$ initial state
\refmeqqttH .

The dominant background for this process are the irreducible backgrounds
\refmettff \refttff,
$$
pp \to gg,q\bar q \to t \bar t \gamma\gamma \tag
$$
here the $q \bar q$ subprocess is of greater
importance than for the signal processes.
This is because we can have the photon radiation off the initial state quark,
which both increases the number of Feynman diagrams and reduces the $Q^2$
flowing through the internal gluon, putting it closer to on mass shell.  At the
SSC the $q \bar q$ background is $20\%$ of the $gg$ background for $m_t=100
\GeV$ increasing to $35\%$ for $m_t=180 \GeV$ \refmeqqttH . At the LHC the $q
\bar q$ background is $50\%$ of the $gg$ background for $m_t = 100 \GeV$
increasing to $120\%$ for $m_t=180 \GeV$ \refmeqqttH. It is clear that the $q
\bar q$ initial state is a very important source of background events,
especially for larger $m_t$, although fortunately for larger $m_t$ the  $t \bar
t \gamma\gamma$  background is far smaller than the $t \bar t H$ signal.

We show the signal and background in \rtable{ttH} and \rfig{ttHcross}. Clearly
we have a good signal to background ratio, with a reasonable number of events.
Especially for large $m_t$ (where the signal doesn't fall as rapidly as the
background, due to the increase in the $ttH$ Yukawa coupling) this gives a high
significance. This requires only the standard luminosity SSC (with
${\cal L}=10^4 \pb^{-1}$), however the LHC still
requires a high luminosity of ${\cal L}=10^5 \pb^{-1}$.

\topinsert
\centerline{\vbox{\offinterlineskip
\hrule
\halign{\vrule#&\strut\quad\hfil$#$\quad
&\vrule\hskip 1pt\vrule#&\quad\hfil$#$\quad
&\vrule#&\quad\hfil$#$\quad
&\vrule#&\quad\hfil$#$\quad
&\vrule\hskip 1pt\vrule#&\quad\hfil$#$\quad
&\vrule#&\quad\hfil$#$\quad
&\vrule#&\quad\hfil$#$\quad
&\vrule\hskip 1pt\vrule#&\quad\hfil$#$\quad
&\vrule#&\quad\hfil$#$\quad
&\vrule#&\quad\hfil$#$\quad&\vrule#\cr
height2pt&\omit&&\multispan5&&\multispan5&&\multispan5&\cr
& m_H\hfil && \multispan5 \hfil $m_t = 100 \GeV$\hfil
&& \multispan5 \hfil $m_t = 140 \GeV$\hfil
&& \multispan5 \hfil $m_t = 180 \GeV$\hfil&\cr
height2pt&\omit&&\multispan5&&\multispan5&&\multispan5&\cr
& (\GeV) && t\bar t H\hfil && t\bar t\gamma\gamma\hfil &&
\omit\hidewidth$\scriptstyle
N_{t\bar t H}\over \scriptstyle \sqrt{N_{t \bar t \gamma\gamma}}$\hidewidth
&& t\bar t H\hfil && t\bar t\gamma\gamma\hfil &&
\omit\hidewidth$\scriptstyle
N_{t\bar t H}\over \scriptstyle \sqrt{N_{t \bar t \gamma\gamma}}$\hidewidth
&& t\bar t H\hfil && t\bar t\gamma\gamma\hfil &&
\omit\hidewidth$\scriptstyle
N_{t\bar t H}\over \scriptstyle \sqrt{N_{t \bar t \gamma\gamma}}$\hidewidth
&\cr
height2pt&\omit&&\omit&&\omit&&\omit&&\omit
&&\omit&&\omit&&\omit&&\omit&&\omit&\cr
\noalign{\hrule}
height2pt&\omit&&\multispan{17}&\cr
&&&\multispan{17}\hfil
Number of events for $\sqrt{s}=16\TeV$ with ${\cal L} = 10^5 \pb^{-1}$
\hfil&\cr
height2pt&\omit&&\multispan{17}&\cr
\noalign{\hrule}
height2pt&\omit&&\omit&&\omit&&\omit&&\omit
&&\omit&&\omit&&\omit&&\omit&&\omit&\cr
& 60 && 18.9 &&  9.0 && 6.3 && 20.9 && 4.0 && 10.5 && 19.8 && 1.7 && 15.2 &\cr
& 70 && 21.3 && 10.4 && 6.6 && 24.7 && 4.7 && 11.4 && 24.2 && 2.0 && 17.1 &\cr
& 80 && 22.6 && 11.3 && 6.7 && 27.1 && 5.0 && 12.1 && 27.4 && 2.1 && 18.9 &\cr
& 90 && 23.2 && 11.6 && 6.8 && 29.0 && 5.0 && 13.0 && 30.3 && 2.2 && 20.4 &\cr
&100 && 23.6 && 11.5 && 7.0 && 29.9 && 4.7 && 13.8 && 32.1 && 2.3 && 21.2 &\cr
&110 && 23.3 && 10.9 && 7.1 && 29.8 && 4.3 && 14.4 && 32.7 && 2.2 && 22.0 &\cr
&120 && 21.8 && 10.0 && 6.9 && 27.7 && 4.0 && 13.9 && 30.9 && 2.1 && 21.3 &\cr
&130 && 18.4 &&  9.0 && 6.1 && 22.9 && 3.8 && 11.7 && 26.0 && 2.0 && 18.4 &\cr
&140 && 13.5 &&  8.2 && 4.7 && 16.3 && 3.7 &&  8.4 && 18.7 && 1.8 && 13.9 &\cr
&150 &&  8.2 &&  7.6 && 3.0 && 10.0 && 3.6 &&  5.3 && 10.9 && 1.8 &&  8.1 &\cr
&160 &&  2.8 &&  7.6 && 1.0 &&  3.1 && 3.3 &&  1.7 &&  3.5 && 1.7 &&  2.8 &\cr
height2pt&\omit&&\omit&&\omit&&\omit&&\omit
&&\omit&&\omit&&\omit&&\omit&&\omit&\cr
\noalign{\hrule}
height2pt&\omit&&\multispan{17}&\cr
&&&\multispan{17}\hfil
Number of events for $\sqrt{s}=40\TeV$ with ${\cal L} = 10^4 \pb^{-1}$
\hfil&\cr
height2pt&\omit&&\multispan{17}&\cr
\noalign{\hrule}
height2pt&\omit&&\omit&&\omit&&\omit&&\omit
&&\omit&&\omit&&\omit&&\omit&&\omit&\cr
& 60 &&  8.9 && 4.4 && 4.2 && 11.0 && 1.7 &&  8.4 && 11.5 && 0.9 && 12.1 &\cr
& 70 && 10.3 && 4.9 && 4.7 && 13.2 && 2.1 &&  9.1 && 14.3 && 1.1 && 13.6 &\cr
& 80 && 11.1 && 5.3 && 4.8 && 14.8 && 2.3 &&  9.8 && 16.4 && 1.2 && 15.0 &\cr
& 90 && 11.8 && 5.4 && 5.1 && 16.1 && 2.4 && 10.4 && 18.4 && 1.2 && 16.8 &\cr
&100 && 12.3 && 5.3 && 5.3 && 16.8 && 2.4 && 10.8 && 19.7 && 1.2 && 18.0 &\cr
&110 && 12.6 && 5.1 && 5.6 && 17.2 && 2.3 && 11.3 && 20.5 && 1.2 && 18.7 &\cr
&120 && 12.1 && 4.9 && 5.5 && 16.2 && 2.2 && 10.9 && 19.7 && 1.2 && 18.0 &\cr
&130 && 10.6 && 4.7 && 4.9 && 13.9 && 2.1 &&  9.6 && 16.8 && 1.2 && 15.3 &\cr
&140 &&  7.9 && 4.4 && 3.8 && 10.1 && 2.1 &&  7.0 && 12.3 && 1.1 && 11.7 &\cr
&150 &&  4.9 && 4.1 && 2.4 &&  6.1 && 2.2 &&  4.1 &&  7.4 && 1.1 &&  7.1 &\cr
&160 &&  1.7 && 3.6 && 0.9 &&  2.0 && 2.3 &&  1.3 &&  2.4 && 0.8 &&  2.9 &\cr
height2pt&\omit&&\omit&&\omit&&\omit&&\omit
&&\omit&&\omit&&\omit&&\omit&&\omit&\cr
}
\hrule}}
\table{ttH}{The numbers of events for the $t\bar t H$ signal, and
$t\bar t \gamma \gamma$ background, assuming a $M_{\gamma\gamma}$
resolution of 3\%; at the high luminosity LHC (${\cal
L}=10^5\pb^{-1}$) and the standard luminosity SSC (${\cal
L}=10^4\pb^{-1}$). We also give the significance, $\scriptstyle
N_{t\bar t H}\over \scriptstyle \sqrt{N_{t \bar t \gamma\gamma}}$,
that this signal would have if it were observed.}
\endinsert
\topinsert
\centerline{\epsfxsize=9 cm\epsfbox{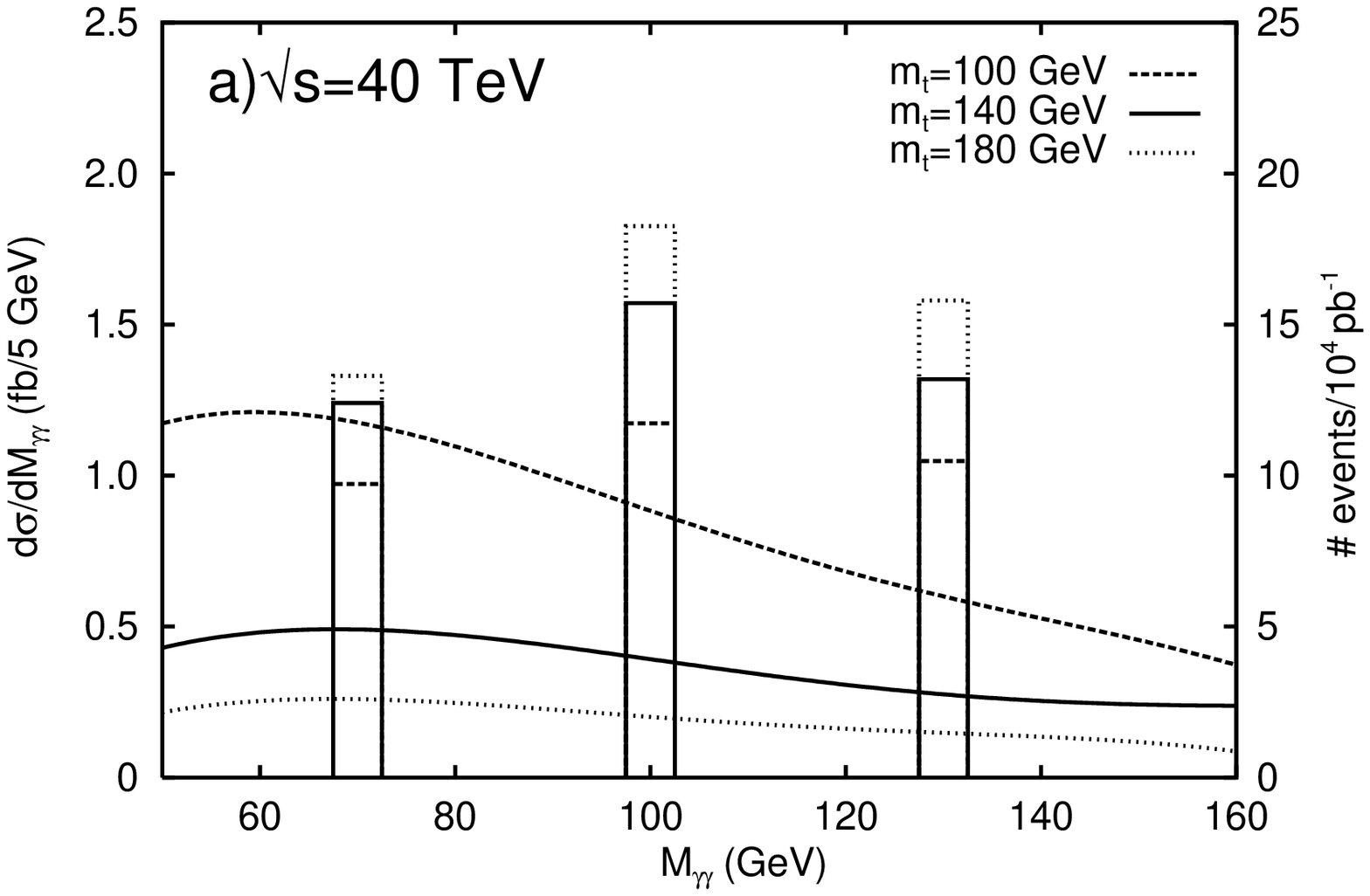}
\qquad\epsfxsize=9 cm\epsfbox{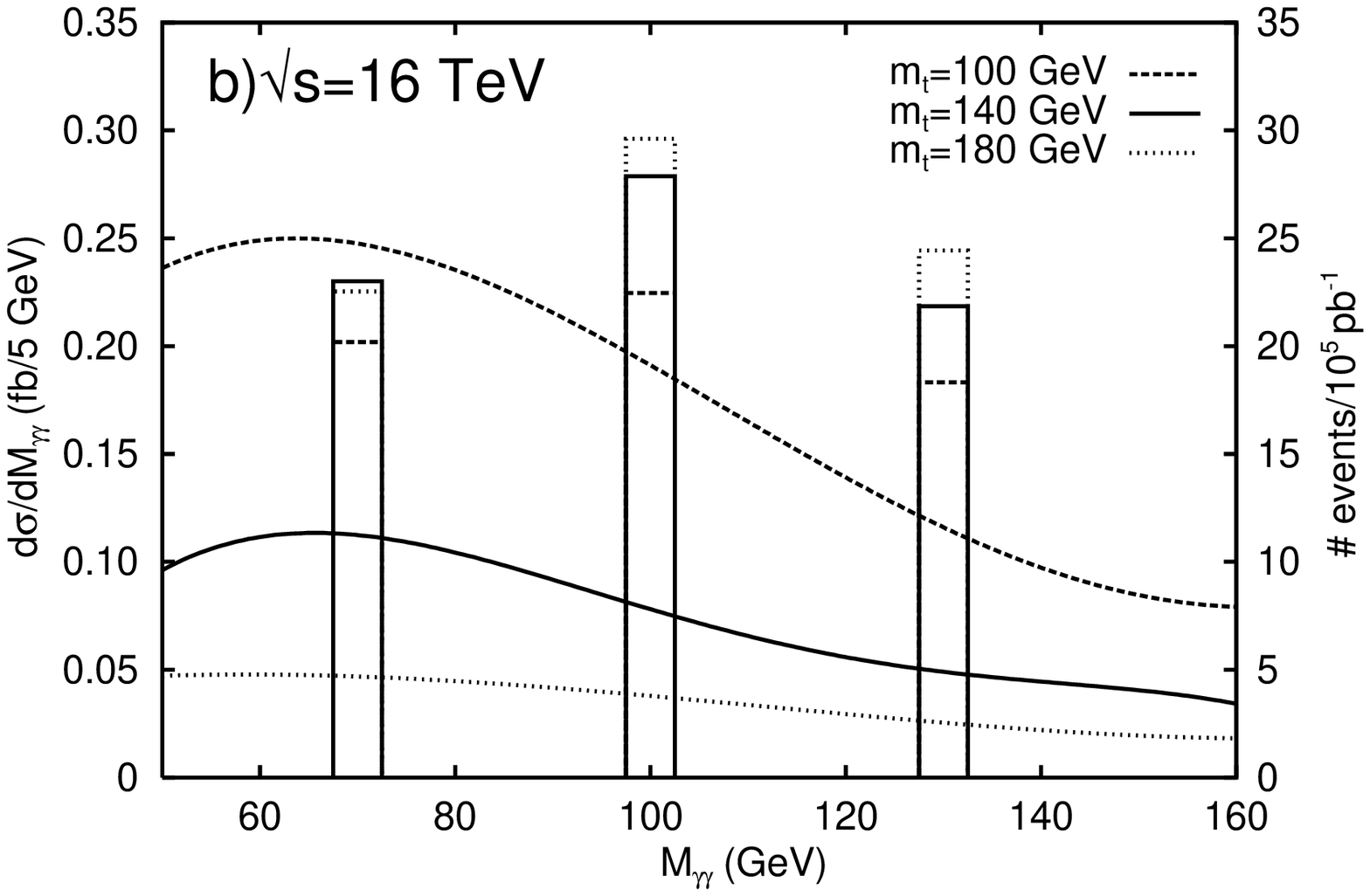}}
\fig{ttHcross}{The differential cross-sections
($d\sigma/dM_{\gamma\gamma}\,(\fb/5\GeV))$ for the process  $pp \to gg,q\bar q
\to t \bar t \gamma \gamma$.
Also shown are the expected numbers of events for the standard luminosity SSC
(${\cal L}=10^4 \pb^{-1}$) and the high luminosity LHC (${\cal L}=10^5
\pb^{-1}$). Superimposed are the cross-sections for the process  $pp \to gg,
q\bar q \to t \bar t H$ for three values of $M_H = 70,100,130 \GeV$.}

\endinsert
\head{Difficulties with NLO corrections ?}

Normally in any process we expect the next--to--leading order (NLO) corrections
to be ${\cal O}(\alpha)$, and so for QCD processes at high energies to only be
of ${\cal O}(10\%)$. However if we consider the process,
$$
q \bar q ' \to W \gamma \qquad ,\tag
$$
then the NLO corrections enhance the cross-section by a factor of about 3
\refNLOWf . Clearly something odd is going on. In this case it is well
understood what is going on, at LO when the photon makes a certain angle  with
respect to the  $q \bar q '$ pair there is a {\it radiation zero} where the
cross-section vanishes identically \refradzero; this is due to destructive
interference between photons radiated off the $W$ and quark legs. This
radiation zero dominates the whole production pattern of the $W\gamma$ pair,
and effectively suppresses radiation over a large range of angles; and this in
turn lowers the total cross-section over the naive expectation. Now when we
move to NLO this radiation zero is absent, and the NLO corrections are ${\cal
O}(\alpha_S)$ of our naive expectation for the $W\gamma$ cross-section; rather
than  ${\cal O}(\alpha_S)$ of the radiation zero dominated correction. This
does not necessarily mean that our perturbation
theory is out of control because as we
more to higher orders in $\alpha_s$ we do not expect similar enhancements; we
should look at this result as saying that the LO cross-section is abnormally
small, rather than the NLO cross-section being abnormally large.

However this raises the question as to how large the NLO corrections are to
$W\gamma\gamma$ production. Unfortunately these have not yet been calculated,
and indeed due to the presence of pentagon diagram in the virtual Feynman
diagrams this calculation stretches current technology to its limit. However we
can ask whether we expect large NLO enhancements in analogy with $W\gamma$
production. Although we only expect the process $q \bar q' \to W \gamma \gamma$
to have radiation zeros when the photons
are parallel \refradzero\refradzeroII, \ie when
$M_{\gamma\gamma}=0$, the matrix element still has very large cancellations
between different Feynman diagrams even for nonzero $M_{\gamma\gamma}$. So in
analogy with $W\gamma$ production we should expect the LO calculation of
$W\gamma\gamma$ production to be abnormally small.

Recently the real contributions to the NLO calculation of
$pp \to W \gamma \gamma$,
$$
\eqalign{
pp &\to q \bar q' \to W \gamma\gamma g \cr
pp &\to g \bar q' \to W \gamma\gamma \bar q\cr
pp &\to q g \to W \gamma\gamma q' \qquad ,\cr
}\tag
$$
have been calculated \refWffj, and here they find that LO
$W\gamma\gamma + 1 {\it jet}$ cross-section is about 3.5 times larger
than the LO $W\gamma\gamma$ cross-section. This result should be taken
with a pinch of salt, for the LO calculation for $W\gamma\gamma + 1 {\it jet}$
production is formally divergent when the final jet becomes collinear
to the incoming partons, or becomes soft; now this divergence cancels
with the 1 loop contribution to $W\gamma\gamma$ production, because when
the final jet is either collinear or soft it can not be experimentally
differentiated from $W\gamma\gamma$ production. This means that until
these virtual 1 loop contributions have been calculated we can not know
how much of the divergent result cancels. In practice what is done is
that a minimum $p_\perp$ cut is imposed on the jet, which at least in
principle keeps the jet experimentally observable, and so separates
the $W\gamma\gamma + 1 {\it jet}$ process from the $W\gamma\gamma$
process. If the cut chosen is too small then we {\it feel} too much
of the divergent cross-section, unfortunately we can not know how small
too small is without doing the complete NLO corrections. Hence we
should be wary of the results of Ref.\refWffj. Having said all this
though Ref.\refWffj\ finds a considerable contribution from
$W\gamma\gamma + 1 { \it jet}$ even with large $p_\perp$ cuts
(\eg $p_\perp > 50\GeV$); this, in conjunction to the apparent
large suppression in the LO $W\gamma\gamma$ production, {\it should}
fill us with worry.

If the NLO corrections to $W\gamma\gamma$ production do turn out to be large
then in principle we can look in the more exclusive channel;
$$
pp \to W\gamma\gamma + 0 \hbox{ \it jets} \tag Wff0j
$$
this will have far smaller corrections as most of the NLO corrections to
$W\gamma\gamma$ production will come from an extra jet produced in the final
state \refNLOWf. However this has a caveat: {\it What is a jet ?}
Or how hard does a
jet have to be before it becomes experimentally observable? Each event has an
underlying event from the breakup of the proton remnants; further fixed order
perturbation theory does not describe low $p_\perp$ jets well. This means that
we can only describe jets well when their $p_\perp$ is larger than some
value. If this value turns out to be large, say 100\GeV,
which seems not unlikely at the LHC and SSC,
then we will still include much of the
NLO cross-section in measuring  the exclusive process \(Wff0j).
At the LHC and SSC we expect
multiple interactions at each bunch crossing, in such an environment it is not
clear what it means to ask to see a process with 0 jets.

\head{Conclusions}

The CERN machines \LEPI/ and \LEPII/ are sensitive to light Higgs bosons with
$M_H \lsim 80 \GeV$; however due to kinematic limits we are
unlikely to be able to probe heavier Higgs bosons. The hadron super colliders
the SSC and the LHC will be sensitive to more massive Higgs bosons with
$M_H \gsim 130 \GeV$. This leaves the mass range
$80 \GeV \lsim M_H \lsim 130 \GeV$ as possibly problematic.

Recently an alternative to looking for the Higgs in isolation at hadron
colliders has been suggested, where we look for the Higgs produced in
association with other massive particles. Although this extra massive particle
decreases the cross-section; it increases the signal to background ratios.
Associated $ZH$ production has too low an
event rate to produce a usable signal.
Associated $WH$ production, at first sight looks far
more promising with 6 times the cross-section of $ZH$ production.
There is a clean signal when the $W$ decays
leptonically to an isolated electron or muon, and the Higgs decays into two
isolated photons. However the major background of $W\gamma\gamma$ production
has only been calculated to LO; and we have good reasons for believing there to
be large NLO corrections. If, in analogy with $W\gamma$ production, the
$W\gamma\gamma$ cross-section is increased by a factor of 3 this will decrease
the significance of our signal by a factor of $\sqrt 3$, alternative to
achieve the same significance we will require 3 times the integrated
luminosity. Nontheless after several years of running at a high luminosity
LHC or SSC (with ${\cal L}=10^5 \pb^{-1}$) this process should give a usable
signal.

Associated $t\bar t H$ production also has a
clean signal in the isolated lepton
(from a $t$ decay) and two photon (from the Higgs decay) mode.
This process has a similar cross-section to $WH$ production at the LHC, but
about
double the $WH$ cross-section at the SSC (due to the rapid growth in the gluon
distribution within the proton at small Bjorken $x$). The main background from
$t \bar t \gamma \gamma$ production does not present a problem, especially for
larger $m_t$; where the background falls due to a more restrictive phase space,
but the signal remains almost constant due to the growth in the $ttH$ Yukawa
coupling. Although the tag for the $t \bar t H$ signal is the same as for $WH$
production (an isolated lepton and two isolated photons), the two types of
event look considerably different. In the $t \bar t H$ case we expect a lot of
jet activity coming from the $t$ quark decays; and this should make the two
types of event easily separable. Indeed if the NLO correction to
$W\gamma\gamma$ production do turn out to be large we may need to insist that
we see a certain amount of hadronic activity in
the event in order to remove this background.
This method gives a signal of reasonable significance at the standard
luminosity
SSC (with ${\cal L}=10^4 \pb^{-1}$) and high luminosity LHC
(with ${\cal L}=10^5 \pb^{-1}$).

Associated production of a Higgs with a heavy particle looks like a very
promising method for searching for a light intermediate mass Higgs boson, with
$80\GeV \lsim M_H \lsim 140 \GeV$ at the SSC and LHC. However there is still
much work to be done in this area before this search technique becomes `gold
plated'. In particular we expect large NLO corrections to the $W\gamma\gamma$
background, which would significantly decrease the significance of the $WH$
signal.

\head{Acknowledgements}

I would like to thank James Stirling and Nigel Glover for helpful
conversations, and SERC for financial support in the form of a research
studentship.

This research was supported in part by the University of Wisconsin Research
Committee with funds granted by the Wisconsin Alumni Research Foundation, in
part by the U.S.~Department of Energy under Contract No.~DE-AC02-76ER00881,
and in part by the Texas National Research Laboratory Commission under
Grant No.~RGFY93-221.

\head{References}

\item\refradmt{U.Amaldi \etal/, {\it Phys. Rev.} {\bf D36} (1987) 1385.}
\item\refHHG{J.F.Gunion, H.E.Haber, G.Kane, and S.Dawson,
      ``Higgs Hunter's Guide'', Addison -- Wesley (1990).}
\item\refHmass{A.A.Carter, Institute of Physics Conference on Nuclear
      and Particle Physics, Glasgow, March 1993.}
\item\refWu{S.L.Wu \etal/, ECFA workshop on LEP200,
      eds. A.B\"ohm and W.Hoogland, Vol II (1987) 312.}
\item\refHZZ{M.Della Nega \etal/, Proceedings of the Large Hadron Collider
      Workshop, Aachen, CERN 90-10 Vol. II (1990) 509,\hfill\break
      J.F.Gunion, G.L.Kane and J. Wudka, {\it Nucl. Phys.} {\bf B299}
      (1988) 239.}
\item\refHBr{Z.Kunszt, and W.J.Stirling, Proceedings of the Large Hadron
      Collider Workshop, Aachen, CERN 90-10 Vol.II (1990) 428,\hfill\break
      D.J.Summers Ph.D. Thesis, University of Durham, unpublished.}
\item\refHff{C.Seez \etal/, Proceedings of the Large Hadron Collider
      Workshop, Aachen, CERN 90-10 Vol.II (1990) 474,\hfil\break
      Proceedings of 1988 Snowmass Workshop on
      ``Physics in the 1990's'', ed. S. Jensen, World Scientific (1989).}
\item\refffake{D.Froidevaux, Proceedings of the Large Hadron
      Collider Workshop, Aachen, CERN 90-10 Vol.II (1990) 444.}
\item\refWH{R.Kleiss, Z.Kunszt, and W.J.Stirling, {\it Phys.Lett.}
      {\bf B253} (1991) 269.}
\item\refttH{W.J. Marciano, and F.E. Paige, {\it Phys.Rev.Lett.}
      {\bf 66} (1991) 2433,\hfil\break
      J.F. Gunion, {\it Phys.Lett.} {\bf B261} (1991) 510.}
\item\refttHbb{J.Dai, J.F.Gunion, and R.Vega,
      University of California - Davis preprint, UCD-93-18.}
\item\refMRS{A.D. Martin, R.G. Roberts, and W.J. Stirling,
       {\it Phys.Lett.} {\bf B306} (1993) 145,\hfill\break
       E. {\it Phys.Lett.} {\bf B309} (1993) 492.}
\item\refjetfake{J.Ohnemus, and W.J.Stirling, {\it Phys. Rev.} {\bf D47}
      (1992) 336.}
\item\refmettff{D.J.Summers, {\it Phys.Lett.} {\bf B277} (1992) 366.}
\item\refttff{Z.Kunszt, Z.Trocsanyi, and W.J.Stirling, {\it Phys.
         Lett.} {\bf B271} (1991) 247,\hfil\break
       A.Ballestrero, and E.Maina, {\it Phys. Lett.}
       {\bf B268} (1991) 437.}
\item\refmeqqttH{D.J.Summers, {\it Phys.Lett.} {\bf B306} (1993) 129.}
\item\refNLOWf{J.Smith, D.Thomas, and W.L.van Neervern, {\it Z.Phys.}
       {\bf C44} (1989) 267,\hfill\break
       J.Ohnemus, {\it Phys.Rev.} {\bf D47} (1993) 940.}
\item\refradzero{R.W.Brown, K.L.Kowalski, and S.J.Brodsky, {\it Phys. Rev.}
        {\bf D28} (1983) 624.}
\item\refradzeroII{R.W.Brown, M.E.Convery, and M.A.Samuel, SLAC preprint,
       SLAC-PUB-6308.}
\item\refWffj{H.Y. Zhou and Y.P. Kuang, {\it Phys.Rev.} {\bf D47}
      (1993) 3680.}

\end